\documentclass[aps,prd,reprint,floatfix,amsmath,nofootinbib,amssymb,superscriptaddress]{revtex4-2}

\usepackage{graphicx}
\usepackage{xcolor}

\usepackage{bm}
\usepackage{slashed}
\allowdisplaybreaks[4]

\usepackage{booktabs}
\usepackage{multirow}

\usepackage[
  unicode=true,
  bookmarks=true,
  bookmarksopen=true,
  bookmarksnumbered=true,
  colorlinks=true,
  linkcolor=blue,
  citecolor=blue,
  urlcolor=blue
]{hyperref}
\usepackage{bookmark}

\begin{document}

\title{Tensor-polarized gluon distribution \texorpdfstring{$h_{1TT}^{g}$}{h1TTg} in the deuteron: QCD evolution and Drell-Yan phenomenology}

\author{Xiupeng Xie}\affiliation{School of Physics, Southeast University, Nanjing
211189, China}

\author{Zhun Lu}
\email{zhunlu@seu.edu.cn}
\affiliation{School of Physics, Southeast University, Nanjing 211189, China}

\begin{abstract}
We study the QCD evolution of the tensor-polarized double-helicity-flip gluon distribution $h_{1TT}^g$ in the deuteron and its impact on proton-deuteron Drell-Yan production. The nonperturbative input is taken from the spin-1 spectator model and evolved at leading order with the autonomous gluon evolution equation. The uncertainty of the model fit is propagated to the tensor-polarized cross-section difference and the asymmetry $A_{E_{xy}}$ using an ensemble of 100 replicas. Compared with the earlier benchmark estimate based on a longitudinally polarized gluon distribution, the evolved spectator-model input gives smaller cross sections in most of the kinematic region and a much stronger decrease of $|A_{E_{xy}}|$ with $Q^2$. At sufficiently large center-of-mass energy, the asymmetry varies only slowly with $\sqrt{s}$, whereas the low-$\sqrt{s}$ and large-$Q^2$ regions probe the poorly constrained large-$x$ tail of the model. We also evolve the tensor-polarized gluon distribution $f_{1LL}^g$ in the coupled singlet-gluon system and assess its contribution to the reference cross-section sum. Its effect remains moderate in the kinematic region where the spectator-model prediction is under reasonable control.
\end{abstract}

\maketitle

\section{Introduction}

Parton distributions in a spin-$1$ hadron contain polarization structures absent for spin-$1/2$ targets. In addition to the unpolarized and vector-polarized sectors, the spin density matrix contains a symmetric traceless rank-two tensor, giving rise to tensor-polarized PDFs and TMDs~\cite{Hoodbhoy:1988am,Bacchetta:2000jk,Boer:2016xqr,Cotogno:2017puy}. For the deuteron, this structure was first formulated through the deep-inelastic structure functions $b_1$--$b_4$~\cite{Hoodbhoy:1988am} and has since been studied through sum rules, tensor-polarized PDFs, and polarized proton-deuteron reactions~\cite{Close:1990zw,Hino:1998ww,Hino:1999qi,Kumano:2010vz,Kumano:2016ude}. The HERMES measurement of $b_1$ found a nonzero tensor-polarized signal, albeit with sizable uncertainties~\cite{HERMES:2005pon}. Conventional proton-neutron convolution calculations do not readily reproduce both its magnitude and $x$ dependence~\cite{Khan:1991qk,Cosyn:2017fbo}, which has led to studies of mesonic, hidden-color, and other nonstandard components of the deuteron~\cite{Miller:2013hla}. Further measurements of tensor-polarized observables have been proposed at Jefferson Lab, Fermilab, NICA, the LHC fixed-target program, and future electron-ion colliders~\cite{Slifer:2014bda,Maxwell:2018gci,Keller:2022abm,Arbuzov:2020cqg,Aidala:2019pit,Accardi:2012qut,
AbdulKhalek:2021gbh,Anderle:2021wcy}.

A particularly interesting distribution in the leading-twist tensor-polarized gluon sector is the collinear function $h_{1TT}^g(x)$. It correlates the gluon linear polarization with the transverse tensor polarization of the spin-$1$ hadron; in a helicity-amplitude representation, it corresponds to a simultaneous two-unit helicity flip of the gluon and the parent hadron~\cite{Jaffe:1989xy,Boer:2016xqr,Cotogno:2017puy}. Angular-momentum conservation therefore forbids it for a spin-$1/2$ target but allows it for $J\geq1$~\cite{Jaffe:1989xy,Artru:1989zv}. The same distribution is often denoted by $\Delta_T g$ or $\Delta_2 G$ and called ``gluon transversity''. However, unlike quark transversity, it is not a density of transversely polarized gluons. In what follows, we instead use the more explicit term ``tensor-polarized double-helicity-flip gluon distribution''. With the polarization conventions of Refs.~\cite{Kumano:2019igu,Kumano:2020gfk},
\begin{align}
h_{1TT}^g(x)&=-\Delta_T g(x)\,.\label{eq:h1TT-convention}
\end{align}
This relation fixes the sign convention used below.

An isolated spin-$1/2$ nucleon cannot carry this distribution. The deuteron $h_{1TT}^g$ therefore cannot be constructed as an incoherent sum of nucleon distributions. It may instead receive contributions from correlations intrinsic to the spin-$1$ system as a whole, including orbital, coherent, $\Delta\Delta$, hidden-color, and other non-nucleonic effects~\cite{Jaffe:1989xy,Nzar:1992ax,Miller:2013hla}. Relating a nonzero $h_{1TT}^g$ to any particular microscopic component of the deuteron is consequently model dependent. Lattice QCD has provided information on the lowest gluonic double-helicity-flip moments~\cite{Detmold:2016gpy}, and positivity inequalities constrain the allowed magnitude~\cite{Cotogno:2017puy}, but no experimental extraction of $h_{1TT}^g(x)$ is presently available. The gluon TMD correlator was classified first for unpolarized and vector-polarized targets~\cite{Mulders:2000sh} and later for tensor-polarized spin-$1$ hadrons~\cite{Boer:2016xqr,Cotogno:2017puy}. Until direct information on $h_{1TT}^g(x)$ becomes available, model calculations provide a practical way to explore its possible size and $x$ dependence. Spectator models, in particular, have been applied extensively to quark and gluon TMD distributions/fragmentation functions in both the T-even and T-odd sectors~\cite{Jakob:1997wg,Bacchetta:2008af,Meissner:2007rx,Lu:2016vqu,Bacchetta:2020vty,Xie:2022lra,Xie:2024cnv,Bacchetta:2024fci}.

In our previous work, we calculated the leading-twist T-even gluon TMDs of the tensor-polarized deuteron in a spectator model~\cite{Xie:2026uja}. In the model the deuteron couples to an off-shell gluon and an on-shell spin-$1$ spectator through three form factors and employs a continuous spectator-mass spectrum following the gluon-spectator strategy of Ref.~\cite{Bacchetta:2020vty}. The model parameters were determined from a replica fit to the unpolarized deuteron gluon distribution. The resulting $h_{1TT}^g(x,\boldsymbol{k}_T^2)$ is nonzero and satisfies the relevant positivity constraints. Its transverse-momentum integral provides the low-scale collinear input for the present calculation. When calculating the corresponding observables in Drell-Yan process, however, this distribution must be evolved to the hard scale. At leading order, $h_{1TT}^g$ obeys the autonomous evolution equation derived by Artru and Mekhfi~\cite{Artru:1989zv}; because it carries two units of helicity in the crossed channel, it does not mix with quark distributions or the ordinary helicity gluon distribution.

A direct hadronic probe was proposed in finite-$q_T$ proton-deuteron Drell-Yan process, with an unpolarized proton colliding on a tensor-polarized deuteron~\cite{Kumano:2019igu}. At the first nonvanishing order in $\alpha_s$, the $q(\bar q)+g\to\gamma^*+q(\bar q)$ Compton channel generates a cross-section difference proportional to  $h_{1TT}^g$. It can be expressed as the difference between the orthogonal linear-polarization states $E_x$ and $E_y$, or reconstructed from conventional transverse deuteron spin states~\cite{Kumano:2020gfk}; the associated tensor asymmetry is denoted by $A_{E_{xy}}$. Related lepton-scattering observables involve characteristic azimuthal modulations of a tensor-polarized target~\cite{Jaffe:1989xy,Maxwell:2018gci}.

The earlier Drell-Yan analysis examined the dilepton-mass, transverse-momentum, and rapidity dependences~\cite{Kumano:2019igu,Kumano:2022cdj}. Because no dedicated double-helicity-flip gluon distribution was available, it used the longitudinally polarized gluon PDF as a benchmark for $\Delta_Tg$. This provided an estimate of the possible signal size but did not incorporate a dynamical spin-$1$ model, the autonomous evolution of $h_{1TT}^g$, or its model uncertainty. Here we evolve the spectator-model $h_{1TT}^g$ to the Drell-Yan scale and propagate the replica ensemble through the calculation of the tensor-polarized cross-section difference and $A_{E_{xy}}$. We examine the dependence on $y$, $q_T$, $Q^2$, and $\sqrt{s}$, retaining both contributions of the quark-antiquark annihilation channel and Compton channel to the cross-section. As an auxiliary study, we also evolve the model gluon distribution $f_{1LL}^{g/d}$ and isolate its effect by replacing $f_1^{g/d}\to f_1^{g/d}+f_{1LL}^{g/d}/2$ in the deuteron-gluon contribution, without including tensor-polarized quark or antiquark corrections.

The remainder of this paper is organized as follows. Section~\ref{section2} presents the Drell-Yan formalism, Sec.~\ref{section3} describes the model input and QCD evolution, Sec.~\ref{section4} presents the numerical results, and Sec.~\ref{section5} summarizes our conclusions.

\section{Tensor-polarized proton-deuteron Drell-Yan formalism}\label{section2}

We consider the Drell-Yan production of a lepton pair in the collision of an unpolarized proton and a tensor-polarized deuteron,
\begin{align}
p(P_A)+d(P_B) \to \gamma^*(q)+X \to \mu^-(k_1) +\mu^+(k_2)+X\,.\nonumber
\end{align}
Here, $P_A$ and $P_B$ denote the proton and deuteron momenta, respectively, while $q=k_1+k_2$ is the virtual-photon momentum. For definiteness, we consider the dimuon channel; the same formalism applies to other light-lepton final states. We work in the hadronic center-of-mass frame, adopting the single-virtual-photon exchange approximation and collinear factorization.  At nonzero $q_T$, the cross sections considered below are evaluated at the first nonvanishing order, $\mathcal{O}(\alpha^2\alpha_s)$. The factorization and renormalization scales are suppressed in this section and will be specified together with the numerical setup in Sec.~\ref{section3}.

In the center of mass frame the proton and deuteron momenta are denoted as
\begin{align}
P_A^\mu&=\frac{\sqrt{s}}{2}(1,0,0,1)\,,
\notag\\
P_B^\mu&=\frac{\sqrt{s}}{2}(1,0,0,-1)\,,
\end{align}
where $s=(P_A+P_B)^2$. The virtual-photon momentum is parametrized by
\begin{align}
q^\mu&=(M_T\cosh y,q_T\cos\phi,q_T\sin\phi,M_T\sinh y)\,,
\end{align}
with
\begin{align}
M_T&=\sqrt{Q^2+q_T^2}\,,\\
\tau&=\frac{Q^2}{s}\,,\\
x_1&=\frac{M_T}{\sqrt{s}}e^y\,,\\
x_2&=\frac{M_T}{\sqrt{s}}e^{-y}\,.
\end{align}
Here, $Q^2=q^2$, $q_T$ and $\phi$ are the magnitude and azimuthal angle of the dimuon transverse momentum, and $y$ is the dimuon rapidity. Denoting the incoming parton momenta as $p_a=x_aP_A$ and $p_b=x_bP_B$, where $x_a$ and $x_b$ are the longitudinal momentum fractions carried by the incoming partons, the on-shell condition for the unobserved final-state parton gives~\cite{Kumano:2019igu}
\begin{align}
x_b&=\frac{x_ax_2-\tau}{x_a-x_1}\,,\\
x_a^{\min}&=\frac{x_1-\tau}{1-x_2}\,.
\end{align}

At finite transverse momentum, the virtual photon may recoil against an additional final-state parton. The lowest-order contribution is then given by $2\to2$ partonic subprocesses at order $\alpha_s$, rather than by the Born subprocess $q\bar q\to\gamma^\ast$ relevant at $q_T=0$. The on-shell condition for the unobserved recoil parton relates $x_b$ to $x_a$ and determines the lower integration boundary.

The physical integration region satisfies $x_a^{\min}\leq x_a\leq1$ and $0<x_b\leq1$. Thus, a given set of $Q^2$, $q_T$, $y$, and $s$ samples only a restricted range of the deuteron momentum fraction $x_b$. This becomes important when the convolution approaches the large-$x_b$ region, where the spectator-model input is less constrained.

The linear polarization states of the deuteron along the two transverse axes are denoted by $E_x$ and $E_y$. In the convention adopted in Ref.~\cite{Kumano:2019igu}, the two states have the same longitudinal tensor component, $S_{B,LL}=1/2$, while their transverse tensor components have opposite signs: $S_{B,TT}^{xx}=-1$ for $E_x$ and $S_{B,TT}^{xx}=+1$ for $E_y$. The two states are related by a rotation of $\pi/2$ around the beam axis.

The difference between the corresponding cross sections removes the terms that are unchanged under this rotation. In particular, the contributions proportional to the unpolarized distribution $f_1$ and the longitudinal tensor-polarized distribution $f_{1LL}$ cancel because their polarization factors are identical for $E_x$ and $E_y$. The rank-two transverse tensor term proportional to $h_{1TT}^g$ changes sign under the same rotation and is isolated by the difference
\begin{align}
\mathrm{d}\sigma_{E_{xy}}&\equiv\mathrm{d}\sigma(E_x)-\mathrm{d}\sigma(E_y)\,.
\end{align}
Conversely, the sum $E_x+E_y$ eliminates the $h_{1TT}^g$ contribution and retains the combination invariant under the $E_x\leftrightarrow E_y$ rotation, namely $f_1+f_{1LL}/2$.

When applying the hard-scattering result of Ref.~\cite{Kumano:2019igu}, we use the sign convention $h_{1TT}^g=-\Delta_T g$ specified in Eq.~\eqref{eq:h1TT-convention}. The minus sign originates from the opposite ordering adopted for the two orthogonal gluon linear-polarization states and does not represent an additional dynamical assumption.

The contraction of the transverse rank-two polarization tensor with the dilepton transverse momentum produces a characteristic $\cos(2\phi)$ modulation. Here, $\phi$ is the azimuthal angle of the dilepton transverse momentum $\boldsymbol q_T$ measured with respect to the $E_x$ axis. It should not be confused with the decay azimuth angle of an individual lepton in a dilepton rest frame. 
An unweighted integration over the full azimuthal range would eliminate the tensor-polarized cross-section difference. Consequently, the signal must be studied either at fixed $\phi$ or through an appropriate $\cos(2\phi)$-weighted projection. The same $E_x-E_y$ combination can also be reconstructed from conventional deuteron spin states polarized along two transverse axes, offering a more practical route to an experimental measurement~\cite{Kumano:2019igu}.

At nonzero $q_T$, the tensor-polarized cross-section difference starts at order $\alpha_s$ through the Compton subprocess $q(\bar q)+g\to\gamma^\ast+q(\bar q)$. The tensor-polarized gluon originates from the deuteron, while the unpolarized proton supplies a quark or antiquark. Using the convention in Eq.~\eqref{eq:h1TT-convention}, the differential cross-section difference is given by~\cite{Kumano:2019igu}
\begin{align}
\frac{\mathrm{d}\sigma_{E_{xy}}}{\mathrm{d}\tau \, \mathrm{d}q_T^2 \, \mathrm{d}\phi \, \mathrm{d}y}=&\frac{\alpha^2\alpha_sC_Fq_T^2}{6\pi s^3}\cos(2\phi)\notag\\
&\times\int_{x_a^{\min}}^1\mathrm{d}x_a \,\frac{1}{(x_ax_b)^2(x_a-x_1)}\notag\\
&\times\frac{1}{(\tau-x_ax_2)^2}\sum_q e_q^2x_a\notag\\
&\times\left[q_A(x_a)+\bar q_A(x_a)\right]x_bh_{1TT}^g(x_b)\,.\label{eq:sigma-exy}
\end{align}
The factorization-scale arguments of the PDFs and the renormalization-scale argument of $\alpha_s$ are suppressed for compactness.

The cross-section difference in Eq.~\eqref{eq:sigma-exy} is linear in the evolved distribution $h_{1TT}^g$. At fixed $\phi$, its sign and magnitude are governed by the weighted convolution of $h_{1TT}^g(x_b)$ with the unpolarized proton quark and antiquark distributions, rather than by the value of $h_{1TT}^g$ at a single momentum fraction. The factor $q_T^2\cos(2\phi)$ reflects the quadrupole structure associated with the double-helicity-flip gluon distribution.

The tensor asymmetry is defined as
\begin{align}
A_{E_{xy}}&=\frac{\mathrm{d}\sigma(E_x)-\mathrm{d}\sigma(E_y)}{\mathrm{d}\sigma(E_x)+\mathrm{d}\sigma(E_y)}\,.
\end{align}
For phenomenological analysis, we define the baseline reference cross-section sum by setting the tensor-polarized distributions $f_{1LL}(x_b)$ in the denominator to zero:
\begin{align}
\mathrm{d}\sigma_{\Sigma}^{(0)}&\equiv \left[\mathrm{d}\sigma(E_x)+\mathrm{d}\sigma(E_y)\right]_{f_{1LL}=0}\,.
\end{align}
The superscript $(0)$ marks this as the \emph{baseline choice}: the reference cross-section sum is constructed from the unpolarized distributions only, and the possible effect of the tensor-polarized $f_{1LL}$ distributions is quantified separately in Sec.~\ref{section4}.
The corresponding baseline asymmetry is
\begin{align}
A_{E_{xy}}^{(0)}&=\frac{\mathrm{d}\sigma_{E_{xy}}}{\mathrm{d}\sigma_{\Sigma}^{(0)}}\,.
\end{align}
Unless explicitly stated otherwise, the notation $A_{E_{xy}}$ used in the numerical results refers to the baseline quantity $A_{E_{xy}}^{(0)}$.
With this notation, $A_{E_{xy}}$ is the ratio of $\mathrm{d}\sigma_{E_{xy}}$ to $\mathrm{d}\sigma_{\Sigma}^{(0)}$, with both quantities understood to be differential in the same set of external variables.

At the same order in $\alpha_s$, the cross-section sum receives contributions from the annihilation subprocess $q\bar q\to\gamma^\ast g$ as well as the two Compton subprocesses $q(\bar q)g\to\gamma^\ast q(\bar q)$ and $gq(\bar q)\to\gamma^\ast q(\bar q)$. We introduce the shorthand notation
\begin{align}
q_A&=f_1^{q/p}(x_a)\,,&q_B&=f_1^{q/d}(x_b)\,,\notag\\
\bar q_A&=f_1^{\bar q/p}(x_a)\,,&\bar q_B&=f_1^{\bar q/d}(x_b)\,,\notag\\
g_A&=f_1^{g/p}(x_a)\,,&g_B&=f_1^{g/d}(x_b)\,.
\label{eq:pdf-shorthand}
\end{align}
Eq.~\eqref{eq:pdf-shorthand} defines the input of the unpolarized PDF  used for the baseline cross-section sum. Tensor-polarized distributions are not included in the main cross-section calculation.

Within this approximation, the differential reference cross-section sum is
\begin{align}
\frac{\mathrm{d}\sigma_{\Sigma}^{(0)}}{\mathrm{d}\tau\,\mathrm{d}q_T^2\,\mathrm{d}\phi\,\mathrm{d}y}=&\frac{\alpha^2\alpha_sC_F}{2\pi\tau s^2}\int_{x_a^{\min}}^1\mathrm{d}x_a \, \frac{1}{(x_a-x_1)x_a^2x_b^2}\sum_q e_q^2\notag\\
&\times\Bigg\{\frac{4}{9}(q_A\bar q_B+\bar q_Aq_B)\mathcal{C}_{q\bar q}\notag\\
&\qquad+\frac{1}{6}(q_A+\bar q_A)g_B\mathcal{C}_{qg}\notag\\
&\qquad+\frac{1}{6}g_A(q_B+\bar q_B)\mathcal{C}_{gq}\Bigg\}\,.
\label{eq:sigma-sum}
\end{align}

The coefficient functions are
\begin{align}
\mathcal{C}_{q\bar q}=&\frac{2\tau(\tau+2x_ax_b-x_1x_b-x_2x_a)}{(x_a-x_1)(x_b-x_2)}\notag\\
&+\frac{x_b^2(x_a-x_1)^2+x_a^2(x_b-x_2)^2}{(x_a-x_1)(x_b-x_2)}\,,\\
\mathcal{C}_{qg}=&\frac{2\tau(\tau-x_1x_b)+x_b^2\left[(x_a-x_1)^2+x_a^2\right]}{x_b(x_a-x_1)}\,,\\
\mathcal{C}_{gq}=&\frac{2\tau(\tau-x_2x_a)+x_a^2\left[(x_b-x_2)^2+x_b^2\right]}{x_a(x_b-x_2)}\,.
\end{align}
The three terms in Eq.~\eqref{eq:sigma-sum} correspond to the $q\bar q$, $qg$, and $gq$ channels, respectively. Retaining all three channels is necessary because their relative contributions vary with the external kinematics and with the momentum fractions sampled on the proton and deuteron sides. In particular, the reference cross-section sum depends not only on the deuteron gluon distribution but also on its quark and antiquark distributions. The additional channels therefore dilute the direct sensitivity of $A_{E_{xy}}$ to $h_{1TT}^g$ and can modify its dependence on $Q^2$, $q_T$, $y$, and $s$.

The expressions above constitute a fixed-order collinear calculation at nonzero $q_T$. In the region $q_T\ll Q$, potentially large logarithms of $Q^2/q_T^2$ and intrinsic transverse-momentum effects may become important and would require a resummed or TMD description. The results obtained very close to $q_T=0$ should therefore be interpreted as fixed-order estimates rather than as precision predictions. In addition, the external kinematics cause the allowed convolution region to extend into large-$x_b$ region, where the spectator-model distributions are less constrained. The corresponding model dependence will be examined explicitly in Secs.~\ref{section3} and~\ref{section4}.

\section{Gluon distribution input and QCD evolution}\label{section3}

The tensor-polarized cross-section difference in Drell-Yan requires the deuteron distribution $h_{1TT}^g$ at the hard scale. 
We take its nonperturbative input from our previous spectator-model calculation and evolve it from the model scale to the relevant scale $Q$. We also evolve the model prediction for $f_{1LL}^g$, which is used only to assess its possible effect on the reference cross-section sum.

The nonperturbative inputs for $h_{1TT}^g$ and $f_{1LL}^g$ are taken from our previous spectator-model calculation of the leading-twist T-even gluon TMDs of a tensor-polarized deuteron~\cite{Xie:2026uja}. In that framework, the deuteron couples to an off-shell gluon and an on-shell spin-1 spectator through an effective vertex containing three form factors, while a continuous spectral distribution is introduced for the spectator mass. The model parameters were determined by fitting the integrated unpolarized gluon distribution to the nNNPDF1.0 deuteron gluon PDF. We refer the reader to Ref.~\cite{Xie:2026uja} for the explicit vertex structure, spectral function, and analytical expressions for the TMDs.

The collinear distributions used as the evolution inputs are obtained through the unweighted transverse-momentum integrals
\begin{align}
h_{1TT}^g(x,Q_0^2)&=\int_0^\infty\mathrm{d}\bm{k}_T^2\,h_{1TT}^g(x,\bm{k}_T^2;Q_0^2)\,,\\
f_{1LL}^g(x,Q_0^2)&=\int_0^\infty\mathrm{d}\bm{k}_T^2\,f_{1LL}^g(x,\bm{k}_T^2;Q_0^2)\,.
\end{align}

Upon unweighted integration over $\bm{k}_T^2$, only two of the leading-twist tensor-polarized T-even gluon TMDs—$f_{1LL}^g$ and $h_{1TT}^g$—yield nonzero collinear counterparts, while the remaining ones integrate to zero.
These two distributions encode different polarization structures. The function $f_{1LL}^g$ modifies the density of unpolarized gluons in a longitudinally tensor-polarized deuteron, whereas $h_{1TT}^g$ correlates gluon linear polarization with the transverse tensor polarization of the target. Their distinct operator structures therefore lead to different evolution equations.

The model parameters were determined by fitting the  unpolarized gluon PDF to the nNNPDF1.0 deuteron gluon distribution~\cite{AbdulKhalek:2019mzd} at the input scale $Q_0=2~\mathrm{GeV}$.
The fit was performed over the range $0.001<x<1$. The fitted spectator-model result for $f_1^g$ is used only to constrain the model parameters. In the calculation of the Drell-Yan process, the unpolarized deuteron quark, antiquark, and gluon PDFs are taken directly from nNNPDF1.0 and are not replaced by the spectator-model distributions. The corresponding unpolarized proton PDFs are taken from NNPDF4.0~\cite{NNPDF:2021njg}.

The uncertainty of the model fit is represented by 100 bootstrap replicas. Each replica contains one correlated set of vertex and spectral-function parameters and is used consistently for both $h_{1TT}^g$ and $f_{1LL}^g$ throughout the evolution. The resulting 100 predictions are propagated to every Drell-Yan observable. At each kinematic point, the 100 replica predictions are ordered, the lowest and highest 16 predictions are discarded, and the envelope of the remaining 68 predictions is taken as the central $68\%$ uncertainty interval. To illustrate the scale dependence of the distributions themselves, we show the results obtained from the representative replica 65.

Although the unpolarized fit covers a broad range in $x$, the tensor-polarized distributions are spectator-model predictions and are not directly constrained by experimental data. Their large-$x$ behavior is particularly sensitive to the form factors, the spectral function, and cancellations among different vertex contributions. In practice, the model distributions become strongly suppressed and less stable when the convolution probes $x\gtrsim0.2$. The corresponding sensitivity at low $\sqrt{s}$ or large $Q^2$ will be discussed in Sec.~\ref{section4}.

For a generic splitting function $P$ and parton distribution $F$, we adopt the convolution convention
\begin{align}
\left[P\otimes F\right](x,\mu^2)&=\int_x^1\frac{\mathrm{d}z}{z}\,P(z)F\left(\frac{x}{z},\mu^2\right)\,.
\end{align}
The scale $\mu$ denotes the collinear factorization scale. Both tensor-polarized inputs are evolved at leading order, consistently with the perturbative accuracy of the hard-scattering calculation.

The double-helicity-flip distribution $h_{1TT}^g$ satisfies an autonomous evolution equation~\cite{Artru:1989zv},
\begin{align}
\frac{ \partial h_{1TT}^g(x,\mu^2) }{ \partial\ln\mu^2 } &= \frac{\alpha_s(\mu^2)}{2\pi} \left[ P_{gg}^{T,(0)} \otimes h_{1TT}^g \right](x,\mu^2)\,.
\label{eq:h1TT-evolution}
\end{align}
The corresponding leading-order gluon splitting function is
\begin{align}
P_{gg}^{T,(0)}(z)&=2C_A\left[\frac{z}{1-z}\right]_++\frac{\beta_0}{2}\delta(1-z)\,.
\label{eq:h1TT-splitting-function}
\end{align}
Here, $\beta_0=\frac{11C_A-4T_Rn_f}{3}$, $C_A=N_c$, $T_R=1/2$, and $n_f$ is the number of active quark flavors. The plus prescription and the $\delta(1-z)$ term combine the real and virtual contributions in the endpoint region.

The autonomous structure of Eq.~\eqref{eq:h1TT-evolution} follows from the two-unit helicity transfer associated with $h_{1TT}^g$. The distribution does not mix with tensor-polarized quark distributions or with the ordinary helicity gluon distribution. Consequently, no quark distribution acts as a perturbative source for $h_{1TT}^g$; its evolved shape is determined entirely by the gluon input and the kernel in Eq.~\eqref{eq:h1TT-splitting-function}. Moreover, since the ordinary unpolarized gluon kernel contains a $1/z$ contribution that is absent here, its evolution does not generate the same strong small-$x$ enhancement as the unpolarized gluon PDF.

The evolution of $f_{1LL}^g$ has a different structure. The distribution is associated with the same twist-2 gluon operator as the ordinary unpolarized gluon PDF, with the tensor polarization entering through the deuteron matrix element. It therefore belongs to the ordinary flavor-singlet evolution sector. We define the tensor-polarized quark singlet as
\begin{align}
\Sigma_{LL}(x,\mu^2)&=\sum_q\left[f_{1LL}^{q/d}(x,\mu^2)+f_{1LL}^{\bar q/d}(x,\mu^2)\right]\,.
\end{align}
At leading order, the coupled singlet-gluon evolution equation is
\begin{align}
\frac{\partial}{\partial\ln\mu^2}
\begin{pmatrix}
\Sigma_{LL}\\
f_{1LL}^g
\end{pmatrix}
&=
\frac{\alpha_s(\mu^2)}{2\pi}
\begin{pmatrix}
P_{qq}^{(0)}
&
2n_fP_{qg}^{(0)}
\\
P_{gq}^{(0)}
&
P_{gg}^{(0)}
\end{pmatrix}
\notag\\
&\quad\otimes
\begin{pmatrix}
\Sigma_{LL}\\
f_{1LL}^g
\end{pmatrix}\,.
\label{eq:f1LL-singlet-evolution}
\end{align}

The splitting functions in Eq.~\eqref{eq:f1LL-singlet-evolution} are the standard leading-order unpolarized splitting kernels. This follows because tensor polarization modifies the hadronic matrix element but not the ultraviolet renormalization of the underlying twist-2 operators.

At the input scale, all tensor-polarized quark and antiquark distributions are set to zero for each active flavor. Equivalently, the tensor-polarized quark singlet vanishes at the initial scale, whereas the gluon distribution is provided by the spectator model:
\begin{align}
\Sigma_{LL}(x,Q_0^2)
&=
0\,,
\notag\\
f_{1LL}^{g/d}(x,Q_0^2)
&=
f_{1LL,\mathrm{model}}^{g/d}(x,Q_0^2)\,.
\end{align}
Starting from these boundary conditions, we evolve the complete leading-order singlet system. The initially vanishing tensor-polarized quark and antiquark distributions are generated dynamically through $g\to q\bar q$ splitting, and the resulting quark singlet subsequently feeds back into the evolution of $f_{1LL}^g$ through the coupled system. Setting the quark inputs to zero at $Q_0$ therefore specifies a purely gluonic nonperturbative boundary condition, rather than a gluon-only evolution approximation.

The main differential cross sections and asymmetries presented in Sec.~\ref{section4} are calculated with $f_{1LL}=0$ in the reference cross-section sum. The evolved gluon component $f_{1LL}^g$ is retained only for a separate comparison of the results obtained with and without the gluonic tensor-polarized correction. Although the coupled evolution also generates $f_{1LL}^{q/d}$ and $f_{1LL}^{\bar q/d}$, these components are not inserted into that auxiliary Drell-Yan comparison. This choice isolates the effect of the model-predicted gluon distribution and should not be interpreted as a complete calculation of all tensor-polarized contributions to the reference cross section.

The evolution equations are solved numerically with QCDNUM~\cite{Botje:2010ay}. A custom leading-order kernel is implemented for $h_{1TT}^g$, whereas $f_{1LL}$ is evolved with the standard unpolarized singlet-gluon system. The calculation is performed in a variable-flavor-number scheme, with the same $x$ grid, scale grid, running coupling, heavy-flavor thresholds, and interpolation settings used for all replicas. Since QCDNUM evolves momentum densities of the form $xF(x,\mu^2)$, the corresponding conversions between $F$ and $xF$ are implemented consistently when constructing the Drell-Yan convolutions.

The factorization and renormalization scales are chosen as
\begin{align}
\mu_F^2&=\mu_R^2=Q^2\,.
\end{align}
This choice identifies the perturbative scale with the dilepton virtuality and avoids introducing an additional scale hierarchy in the fixed-order calculation. Scale variations and missing higher-order corrections are not included in the uncertainty bands presented below.

\begin{figure}[t]
\centering
\includegraphics[width=\columnwidth]{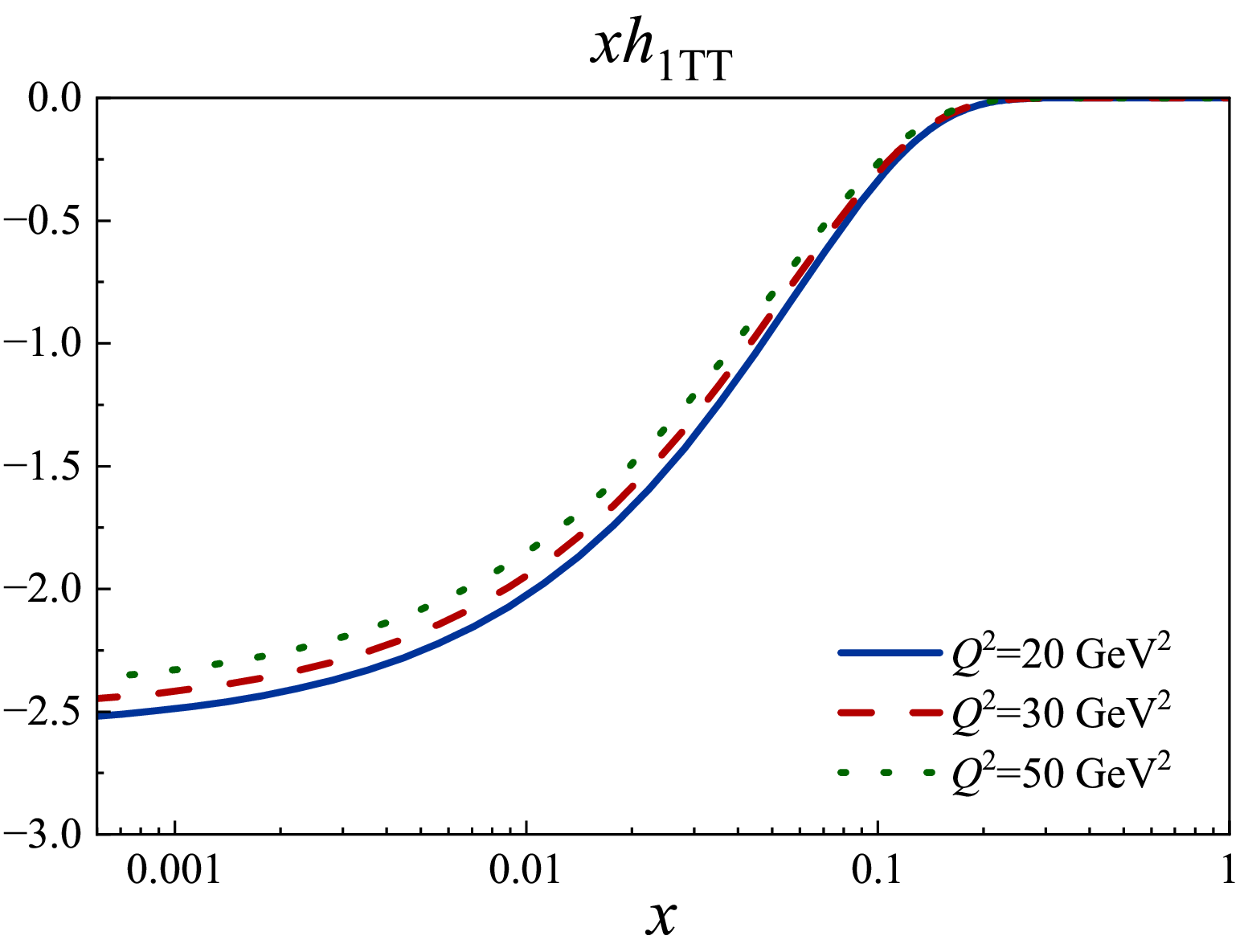}
\caption{Leading-order QCD evolution of $xh_{1TT}^g(x,Q^2)$ from the model input scale $Q_0^2=4~\mathrm{GeV}^2$ to the selected values of $Q^2$ indicated in the figure. Solid line: $Q^2=20~\mathrm{GeV}^2$; dashed line: $Q^2=30~\mathrm{GeV}^2$; dotted line: $Q^2=50~\mathrm{GeV}^2$.}
\label{fig:h1TT-evolution}
\end{figure}

\begin{figure}[t]
\centering
\includegraphics[width=\columnwidth]{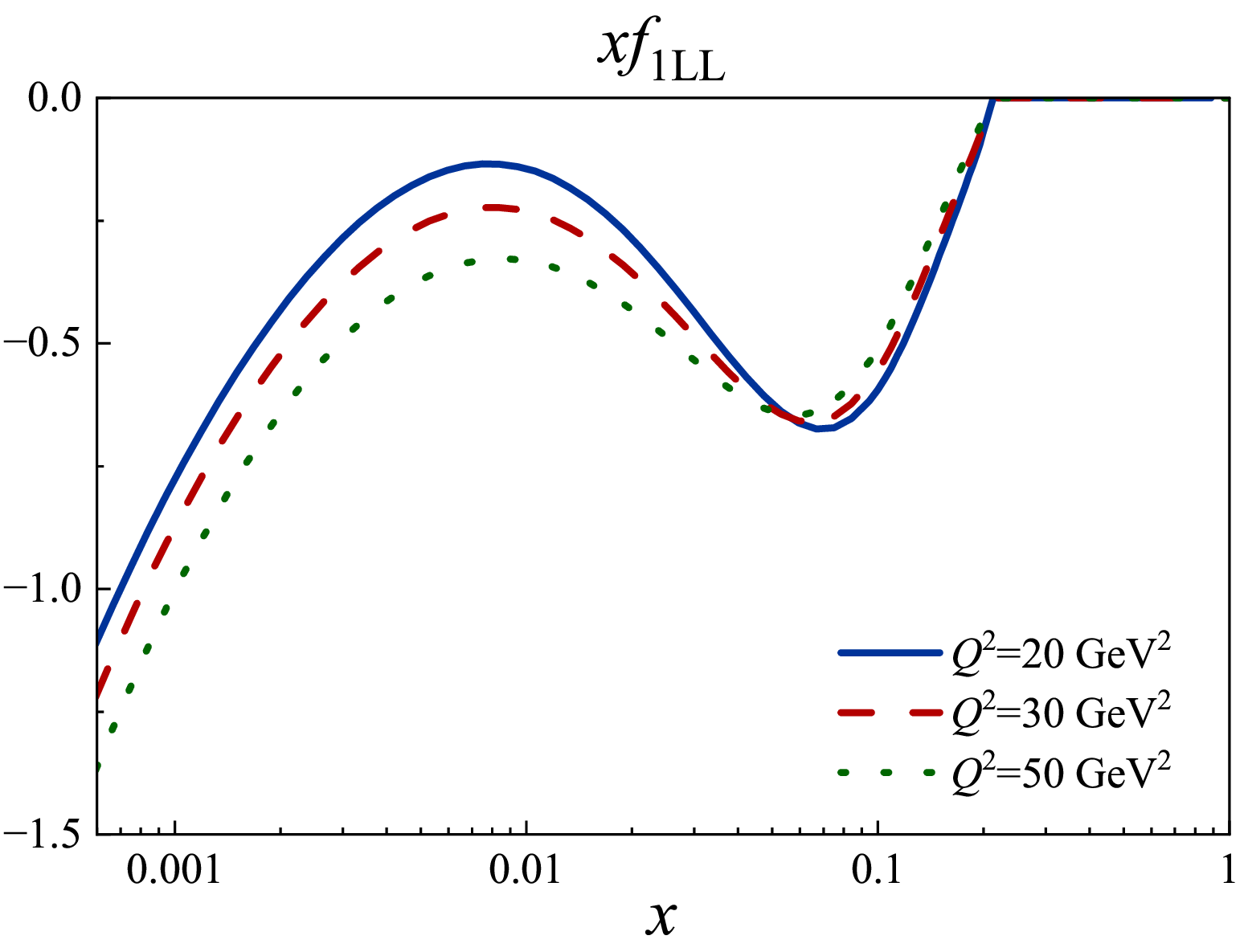}
\caption{Leading-order singlet evolution of $xf_{1LL}^g(x,Q^2)$ from the purely gluonic model input at $Q_0^2=4~\mathrm{GeV}^2$ to the selected values of $Q^2$ indicated in the figure. Solid line: $Q^2=20~\mathrm{GeV}^2$; dashed line: $Q^2=30~\mathrm{GeV}^2$; dotted line: $Q^2=50~\mathrm{GeV}^2$. The tensor-polarized quark and antiquark inputs are set to zero at $Q_0$, but are generated dynamically through the coupled singlet-gluon evolution.}
\label{fig:f1LL-evolution}
\end{figure}

The 100 model replicas are evolved independently from $Q_0^2=4~\mathrm{GeV}^2$ to $Q^2=16,20,30,40,50,60,70,80~\mathrm{GeV}^2$. The same replica number is used for $h_{1TT}^g$ and $f_{1LL}^g$ throughout the evolution. To display the scale dependence without obscuring the characteristic evolution pattern, Figs.~\ref{fig:h1TT-evolution} and~\ref{fig:f1LL-evolution} show the results for the representative replica 65 only. The complete set of evolved replicas is retained for the uncertainty propagation to the Drell-Yan observables discussed in Sec.~\ref{section4}.

Fig.~\ref{fig:h1TT-evolution} shows the scale dependence of the tensor-polarized double-helicity-flip gluon distribution. Evolution redistributes the model input in $x$ and generally suppresses its moderate- and large-$x$ components as the scale increases. Because the evolution is autonomous, the scale dependence and replica uncertainty are determined entirely by the initial gluon distribution and the double-helicity-flip kernel. No quark input or radiatively generated quark distribution contributes to this evolution.

At large $x$, the model distributions are strongly suppressed and become more sensitive to the spectator parameters. In this region, even a small absolute change can produce a sizable relative variation; a broad relative spread should therefore not be interpreted as evidence for a large double-helicity-flip gluon density. 
The numerical result also becomes more sensitive to the $\bm{k}_T$ integration used to construct the model input and, in the Drell-Yan process, to the longitudinal momentum-fraction integration. Cancellations among small contributions can further enhance these relative variations. We therefore treat predictions dominated by the large-$x_b$ region with particular caution.

The evolved gluon component $f_{1LL}^g$ is displayed in Fig.~\ref{fig:f1LL-evolution}. In contrast to $h_{1TT}^g$, its scale dependence is governed by both the ordinary gluon splitting kernel and quark-gluon mixing. The standard gluon kernel contains a small-$z$ enhancement that is absent from the double-helicity-flip kernel, while the coupled evolution transfers part of the gluon input into the tensor-polarized quark singlet. Thus, although the two distributions are obtained from the same spectator-model parameter sets, their scale dependence can be quite different.

Only the evolved gluon curve in Fig.~\ref{fig:f1LL-evolution} will be used in the auxiliary comparison in Sec.~\ref{section4}. It modifies the deuteron gluon contribution to the reference cross-section sum but does not contribute to the $E_x-E_y$ cross-section difference. The baseline cross sections and asymmetries are independent of $f_{1LL}^g$.

Although only replica 65 is displayed in Figs.~\ref{fig:h1TT-evolution} and~\ref{fig:f1LL-evolution}, all 100 replicas are evolved and retained for the subsequent uncertainty analysis. The uncertainty bands for the Drell-Yan observables presented in Sec.~\ref{section4} therefore reflect only the propagated spectator-model parameter uncertainty. They do not include uncertainties from the proton or deuteron unpolarized PDFs, variations of the factorization and renormalization scales, missing higher-order evolution and hard-scattering corrections, or the omission of the radiatively generated tensor-polarized quark distributions from the auxiliary $f_{1LL}$ comparison. The bands shown below should therefore be understood as spectator-model parameter uncertainties only.

\section{Numerical results}\label{section4}

We next turn to the proton-deuteron Drell-Yan observables obtained with the evolved spectator-model $h_{1TT}^g$. Unless stated otherwise, we set $\phi=0$, where the $\cos(2\phi)$ modulation is maximal. For the rapidity, transverse-momentum, invariant-mass, and $Q^2$-dependent asymmetry results, we use
\begin{align}
s&=450~\mathrm{GeV}^2\,,
\end{align}
corresponding to $\sqrt{s}\simeq21.2~\mathrm{GeV}$. When the energy dependence is studied, $s$ is varied with the remaining external kinematics fixed as specified below.

We identify the hard scale with $Q^2=M_{\mu\mu}^2$. The tensor-polarized cross-section difference is negative in the kinematic region considered. We therefore display its absolute magnitude and, following Ref.~\cite{Kumano:2019igu}, also use $|A_{E_{xy}}|$ when comparing with the previous benchmark calculation. The baseline results use unpolarized proton and deuteron PDFs in the reference cross-section sum and omit $f_{1LL}$; its gluonic contribution is examined separately below.

Replica 65 is shown as the representative curve, while the shaded bands denote the central $68\%$ replica intervals defined in Sec.~\ref{section3}. The baseline uncertainty is obtained by propagating the $h_{1TT}^g$ replicas, whereas the auxiliary $f_{1LL}^g$ comparison analogously propagates the $f_{1LL}^g$ replicas. Uncertainties from unpolarized PDFs, scale variations, and missing higher-order corrections are not included. The benchmark curves of Ref.~\cite{Kumano:2019igu}, which used the longitudinally polarized gluon distribution as a proxy for the unknown double-helicity-flip distribution, are shown for comparison.

\begin{figure}[t]
\centering
\includegraphics[width=\columnwidth]{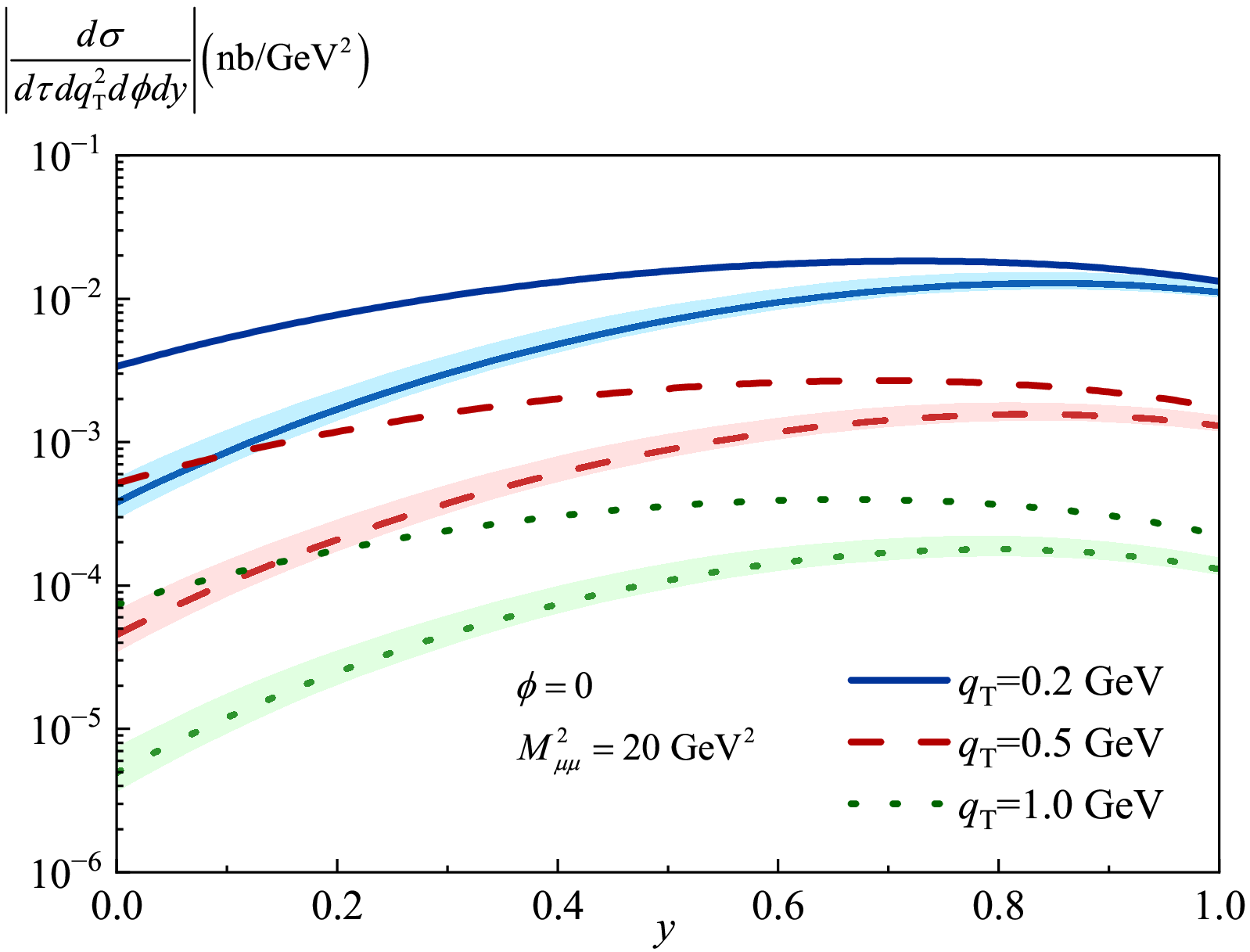}
\caption{Magnitude of the cross-section difference in the tensor-polarized $pd$ Drell-Yan as a function of the dimuon rapidity $y$ at $Q^2=M_{\mu\mu}^2=20~\mathrm{GeV}^2$, $s=450~\mathrm{GeV}^2$, and $\phi=0$. The solid, dashed, and dotted curves correspond to $q_T=0.2$, $0.5$, and $1.0~\mathrm{GeV}$, respectively. The shaded bands show the present spectator-model predictions with the propagated replica uncertainty, while the benchmark curves are taken from Ref.~\cite{Kumano:2019igu}.}
\label{fig:sigma-y}
\end{figure}

Fig.~\ref{fig:sigma-y} shows the rapidity dependence of the tensor-polarized cross-section difference at $Q^2=M_{\mu\mu}^2=20~\mathrm{GeV}^2$. The broad maximum at positive rapidity arises from the interplay between the phase-space boundary and the momentum fractions sampled on the proton and deuteron sides: varying $y$ shifts both $x_a$ and $x_b$, while the decrease toward the edge of the interval reflects the restricted partonic phase space. 
For all three values of $q_T$, the present cross sections are generally smaller than the benchmark results of Ref.~\cite{Kumano:2019igu}. This difference is not simply an overall normalization effect, because the evolved spectator-model $h_{1TT}^g$ and the longitudinally polarized gluon PDF used in the benchmark have different $x$ dependences.

\begin{figure}[t]
\centering
\includegraphics[width=\columnwidth]{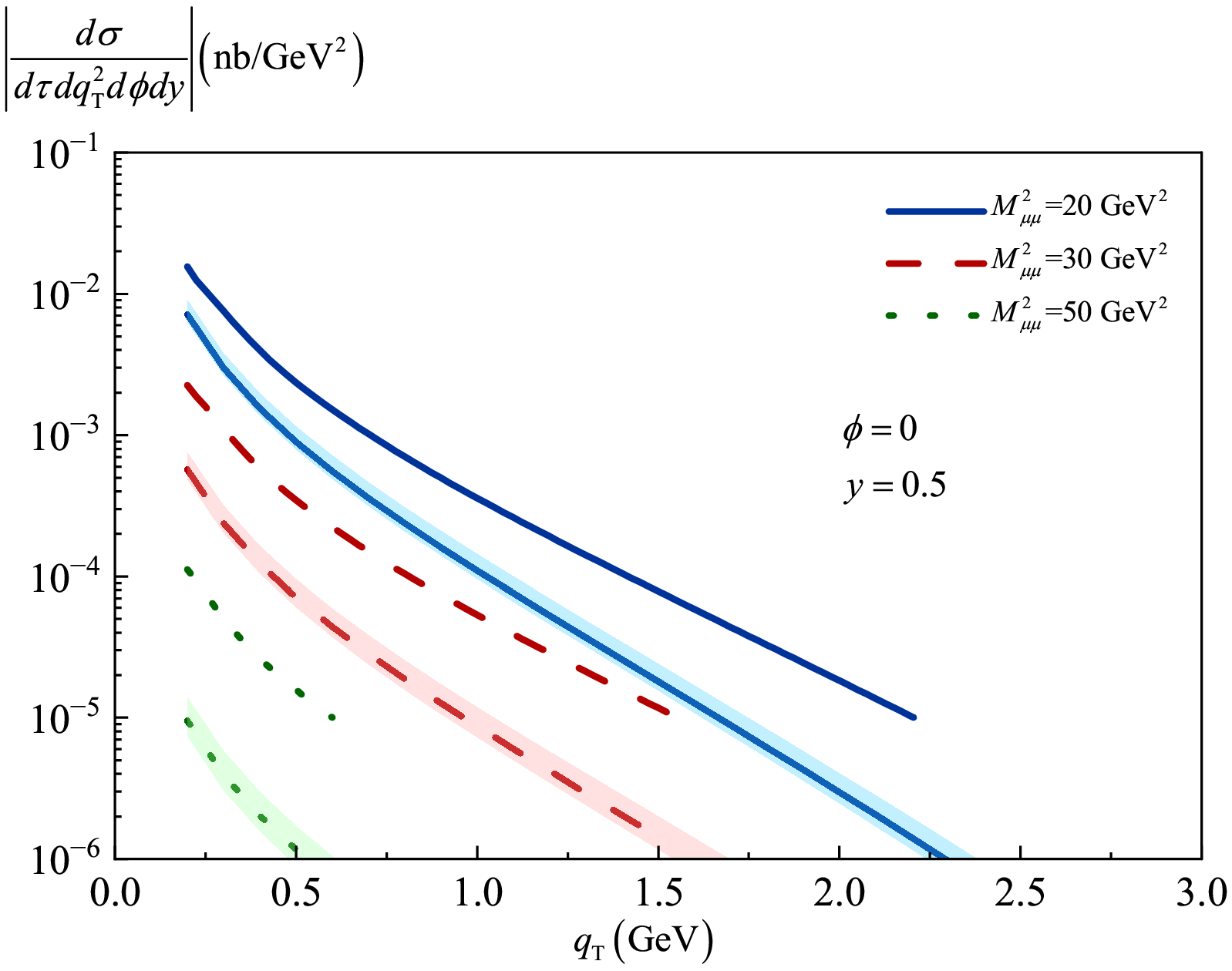}
\caption{Magnitude of the cross-section difference in the tensor-polarized $pd$ Drell-Yan as a function of $q_T$ at $Q^2=M_{\mu\mu}^2=20$, $30$, and $50~\mathrm{GeV}^2$, with $s=450~\mathrm{GeV}^2$, $y=0.5$, and $\phi=0$. The shaded bands show the present results, and the benchmark curves are taken from Ref.~\cite{Kumano:2019igu}.}
\label{fig:sigma-qT}
\end{figure}

Fig.~\ref{fig:sigma-qT} shows the transverse-momentum dependence at $Q^2=20$, $30$, and $50~\mathrm{GeV}^2$. Despite the explicit $q_T^2$ factor in Eq.~\eqref{eq:sigma-exy}, the cross section decreases rapidly as $q_T$ increases. Increasing $q_T$ raises the transverse mass and shifts the lower integration boundary, thereby restricting the available phase space and driving the convolution toward larger momentum fractions; these effects overcome the explicit $q_T^2$ factor, as also found in Ref.~\cite{Kumano:2019igu}. The suppression becomes stronger with increasing $Q^2$, and at $Q^2=50~\mathrm{GeV}^2$ the present result falls below $10^{-5}~\mathrm{nb/GeV}^2$ over most of the displayed range.

The lowest $q_T$ values satisfy $q_T\ll Q$, where logarithms of $Q^2/q_T^2$ can become important. These points should therefore be regarded as fixed-order estimates rather than precision predictions, although they remain useful for comparison with Ref.~\cite{Kumano:2019igu}, which employed the same fixed-order framework.

\begin{figure}[t]
\centering
\includegraphics[width=\columnwidth]{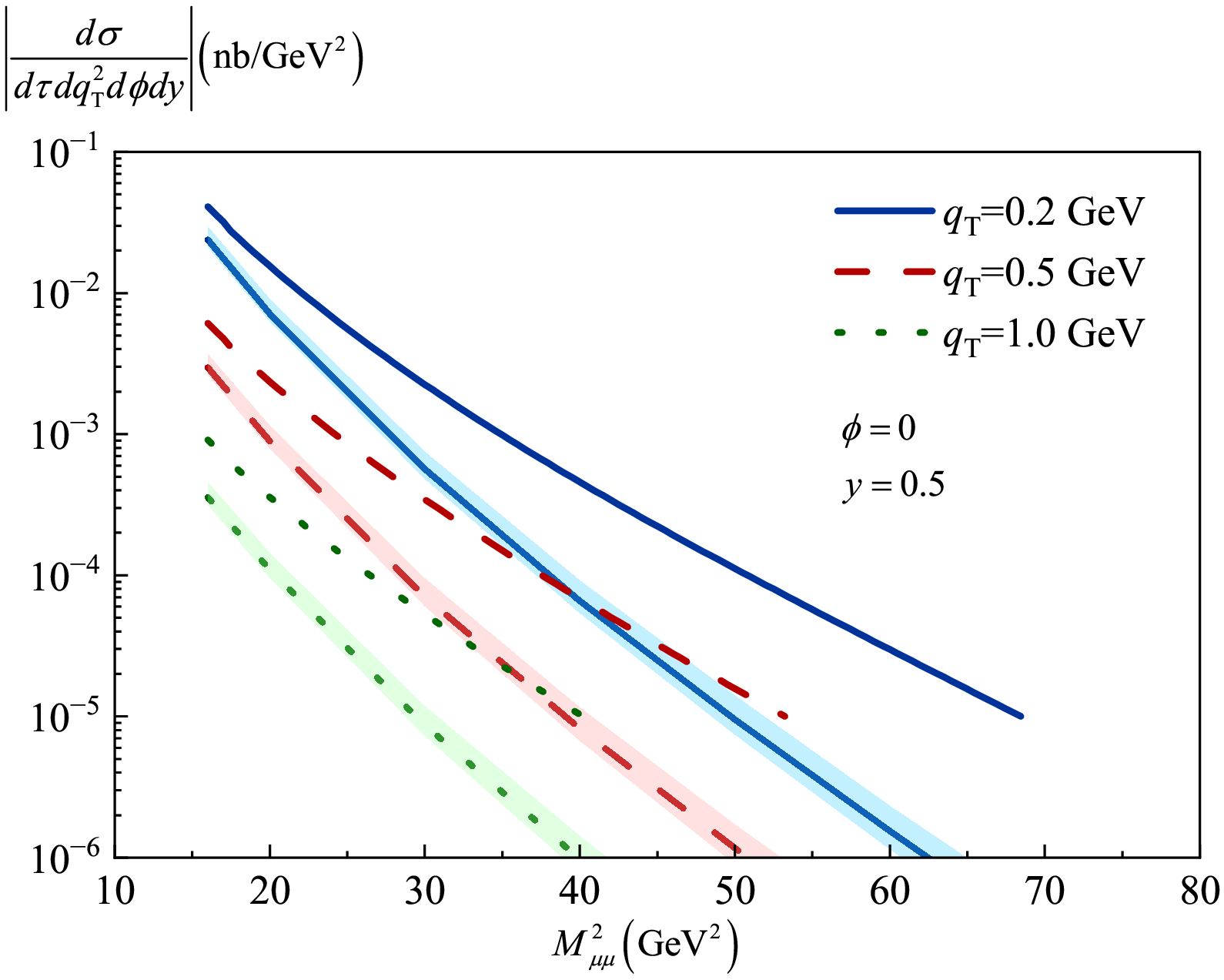}
\caption{Magnitude of the cross-section difference in the tensor-polarized $pd$ Drell-Yan as a function of $Q^2=M_{\mu\mu}^2$ at $q_T=0.2$, $0.5$, and $1.0~\mathrm{GeV}$, with $s=450~\mathrm{GeV}^2$, $y=0.5$, and $\phi=0$. The shaded bands show the present evolved spectator-model results, while the benchmark curves are taken from Ref.~\cite{Kumano:2019igu}.
}
\label{fig:sigma-Q2}
\end{figure}

Fig.~\ref{fig:sigma-Q2} shows the $Q^2$ dependence of the cross-section difference. The rapid decrease comes from three effects acting in the same direction: the available phase space shrinks, $\alpha_s(Q^2)$ decreases, and the evolved $h_{1TT}^g$ is suppressed in the moderate- and large-$x$ region. The falloff is consequently much steeper than in the benchmark calculation, where $\Delta_T g$ was identified with a longitudinally polarized gluon PDF~\cite{Kumano:2019igu}.

At the largest $Q^2$, the convolution is progressively shifted toward the poorly constrained large-$x_b$ tail of the spectator-model distribution. Since the model becomes strongly suppressed and less stable as the integration region extends into $x_b\gtrsim0.2$, the highest-$Q^2$ predictions should be interpreted more cautiously than those at intermediate scales.

\begin{figure}[t]
\centering
\includegraphics[width=\columnwidth]{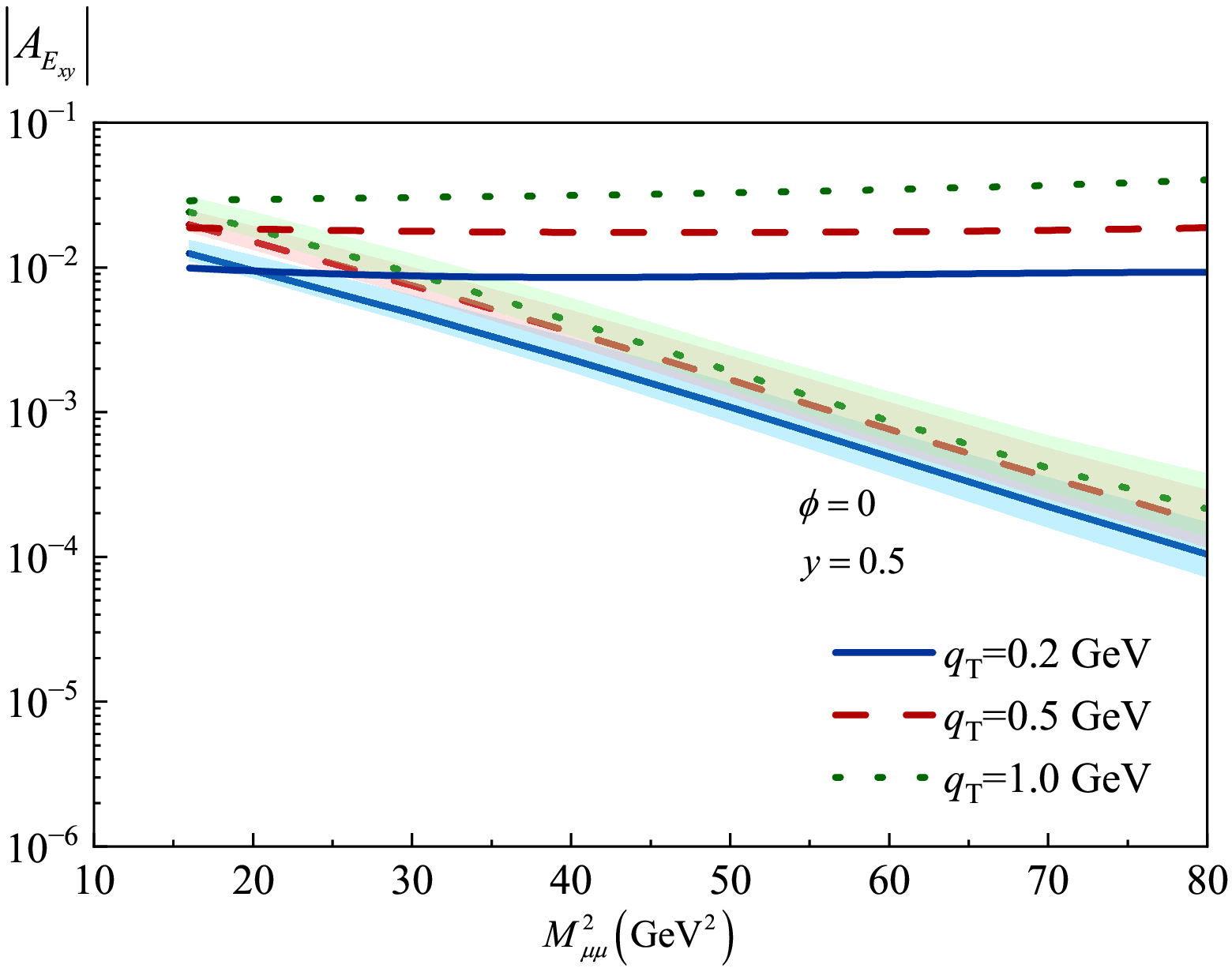}
\caption{The tensor asymmetry $|A_{E_{xy}}|$ as a function of $Q^2=M_{\mu\mu}^2$ at $q_T=0.2$, $0.5$, and $1.0~\mathrm{GeV}$, with $s=450~\mathrm{GeV}^2$, $y=0.5$, and $\phi=0$. The shaded bands are the present predictions without an $f_{1LL}$ correction in the reference cross-section sum. The benchmark curves are taken from Ref.~\cite{Kumano:2019igu}.}
\label{fig:AExy-Q2}
\end{figure}

Fig.~\ref{fig:AExy-Q2} shows the magnitude of the tensor asymmetry. At the lower end of the displayed range, the present results are of order $10^{-2}$, comparable in magnitude to the benchmark estimates, but their scale dependence is markedly different: the benchmark curves remain nearly constant, whereas the present asymmetry decreases to the $10^{-4}$ level toward the upper end of the range.

The decrease does not cancel between the numerator and denominator because the numerator contains the evolved $h_{1TT}^g$, 
whereas the reference cross-section sum in the denominator receives contributions from ordinary quark, antiquark, and gluon PDFs through all three channels in Eq.~\eqref{eq:sigma-sum}. Because these ingredients evolve differently with $Q^2$ and sample different $x_b$ regions, the scale dependence does not cancel in the ratio. The ordering with $q_T$ can also differ from that of the absolute cross section because both the numerator and denominator are suppressed as $q_T$ increases. At large $Q^2$, the increasing sensitivity to the strongly suppressed large-$x_b$ tail again makes the intermediate-$Q^2$ region more robust for quantitative phenomenology.

\begin{figure}[t]
\centering
\includegraphics[width=\columnwidth]{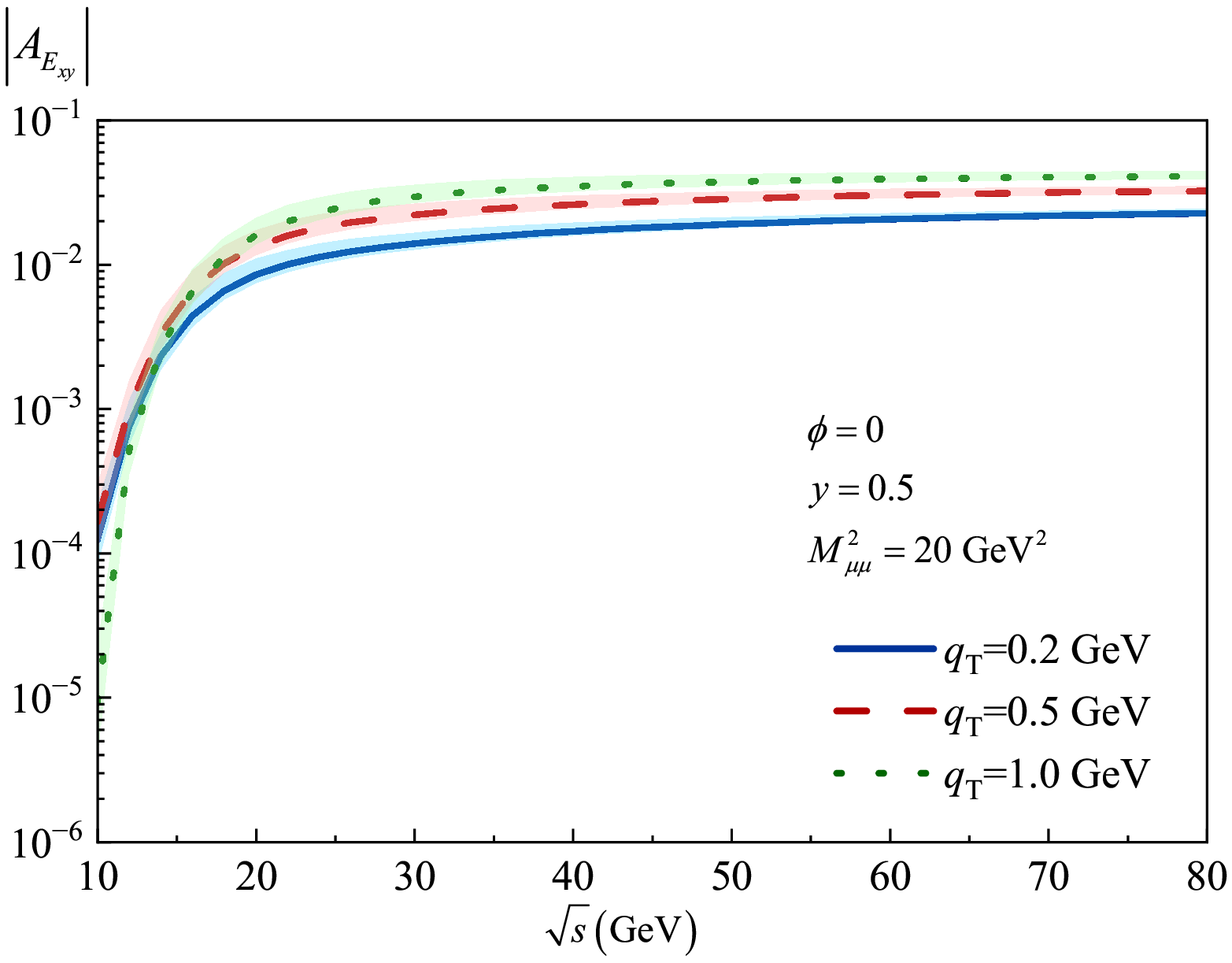}
\caption{The tensor asymmetry $|A_{E_{xy}}|$ as a function of $\sqrt{s}$ for $Q^2=M_{\mu\mu}^2=20~\mathrm{GeV}^2$, $q_T=0.2$, $0.5$, and $1.0~\mathrm{GeV}$, with $y=0.5$ and $\phi=0$. The results correspond to the baseline calculation without the $f_{1LL}$ correction.}
\label{fig:AExy-s}
\end{figure}

The energy dependence at $Q^2=20~\mathrm{GeV}^2$, $y=0.5$, and $\phi=0$ is shown in Fig.~\ref{fig:AExy-s}. The asymmetry rises rapidly away from the low-energy boundary and approaches a slowly varying plateau for $\sqrt{s}\gtrsim30~\mathrm{GeV}$. At sufficiently large $\sqrt{s}$, the numerator and denominator probe a relatively stable small- and intermediate-$x$ region, so that their remaining energy dependences largely cancel in the ratio.

The low-energy region is less reliable. At $\sqrt{s}=15~\mathrm{GeV}$, the minimum accessible deuteron momentum fractions are already $x_{b,\min}=0.181$, $0.184$, and $0.194$ for $q_T=0.2$, $0.5$, and $1.0~\mathrm{GeV}$, respectively, while the integration extends to $x_b=1$. Below roughly $\sqrt{s}=15~\mathrm{GeV}$, the rapid variation and widening replica band are therefore more naturally attributed to the poorly constrained large-$x_b$ tail than to a genuine change in the underlying dynamics. We retain this region to illustrate the onset of this sensitivity, while quantitative discussion is focused on $\sqrt{s}\gtrsim15$--$20~\mathrm{GeV}$.

\begin{figure}[t]
\centering
\includegraphics[width=\columnwidth]{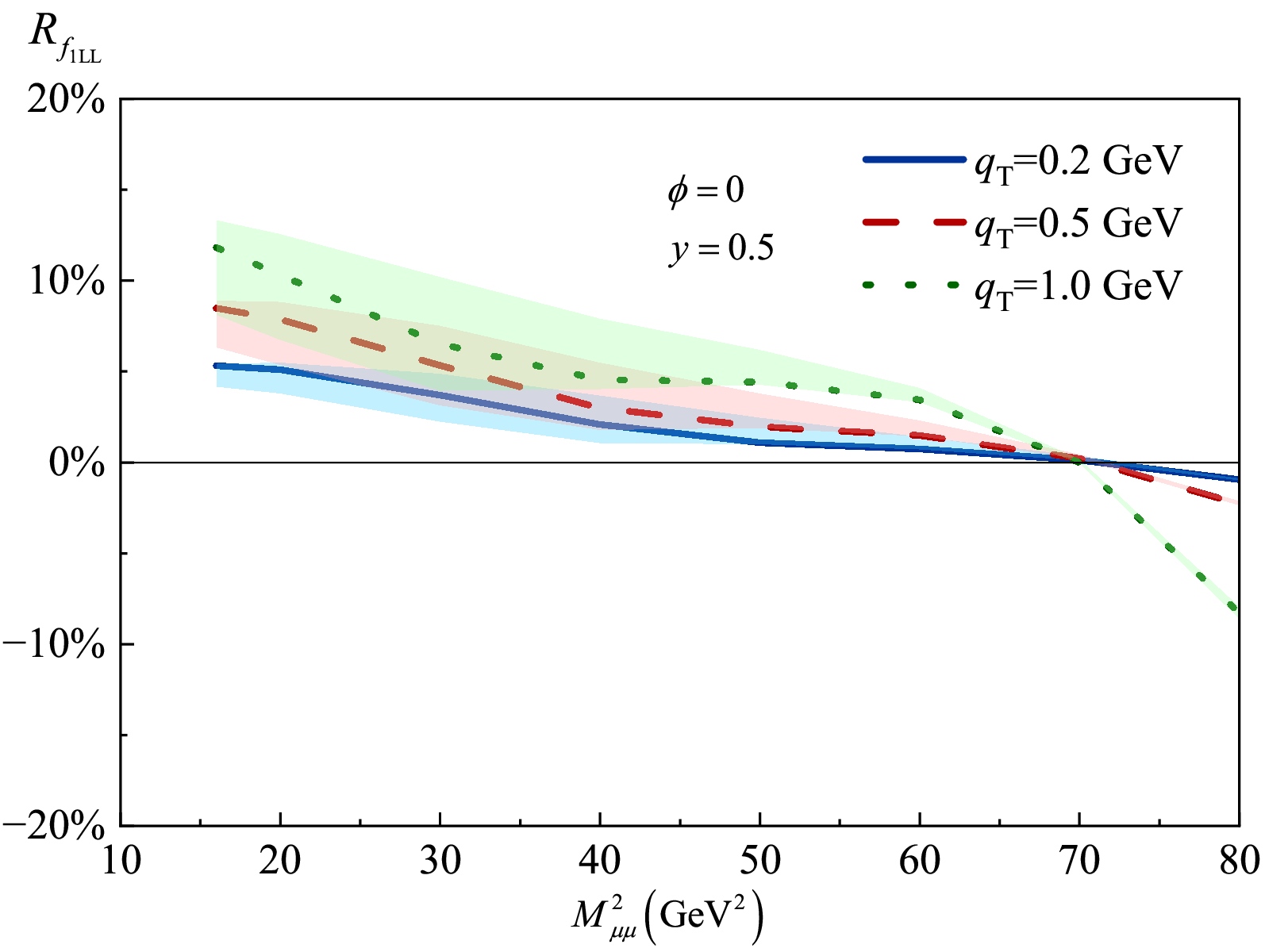}
\caption{Relative change of $A_{E_{xy}}$ induced by including $f_{1LL}^{g/d}/2$ in the deuteron gluon contribution to the reference cross-section sum, shown as a function of $Q^2$ for $q_T=0.2$, $0.5$, and $1.0~\mathrm{GeV}$ at $s=450~\mathrm{GeV}^2$, $y=0.5$, and $\phi=0$.}
\label{fig:f1LL-correction-Q2}
\end{figure}

We finally estimate the correction associated with the evolved gluon distribution $f_{1LL}^{g/d}$ shown in Fig.~\ref{fig:f1LL-evolution}. Only the deuteron gluon distribution in the $q_Ag_B$ Compton channel is modified,
\begin{align}
g_B(x_b,Q^2)&\longrightarrow g_B^{(LL)}(x_b,Q^2)\notag\\
&=f_1^{g/d}(x_b,Q^2)+\frac{1}{2}f_{1LL}^{g/d}(x_b,Q^2)\,,
\label{eq:f1LL-correction}
\end{align}
while all other ingredients are kept unchanged. The tensor-polarized quark and antiquark distributions generated by the singlet evolution are not included, so this comparison isolates the model-predicted gluonic $f_{1LL}^g$ contribution rather than providing a complete tensor-polarized reference cross section.

Since $f_{1LL}^g$ cancels from the $E_x-E_y$ cross-section difference, the numerator of $A_{E_{xy}}$ is unchanged. We denote the baseline asymmetry by $A_{E_{xy}}^{(0)}$ and the asymmetry after the replacement in Eq.~\eqref{eq:f1LL-correction} by $A_{E_{xy}}^{(LL)}$. The relative correction is defined as
\begin{align}
R_{f_{1LL}}&=\frac{A_{E_{xy}}^{(LL)}-A_{E_{xy}}^{(0)}}{A_{E_{xy}}^{(0)}}\times100\%\,.
\end{align}
Because the two asymmetries have the same numerator, the dependence on $h_{1TT}^g$ cancels exactly in $R_{f_{1LL}}$, which therefore measures only the relative modification of the reference cross-section sum induced by the gluonic $f_{1LL}^g$ contribution.

Fig.~\ref{fig:f1LL-correction-Q2} shows the correction as a function of $Q^2$. At $Q^2=16~\mathrm{GeV}^2$, the representative values are approximately $5.4\%$, $8.5\%$, and $11.8\%$ for $q_T=0.2$, $0.5$, and $1.0~\mathrm{GeV}$, respectively. The correction decreases overall with $Q^2$ and is systematically larger at larger $q_T$, reflecting the increasing relative importance of the deuteron-gluon channel.

Around $Q^2=80~\mathrm{GeV}^2$, the correction changes sign. Because $R_{f_{1LL}}$ depends only on the $f_{1LL}^g$ correction to the reference cross-section sum, we do not assign physical significance to this sign change. In this region the convolution probes the strongly suppressed large-$x_b$ tail, where the result becomes increasingly sensitive to the spectator-model parametrization, numerical accuracy, and cancellations among small contributions.

\begin{figure}[t]
\centering
\includegraphics[width=\columnwidth]{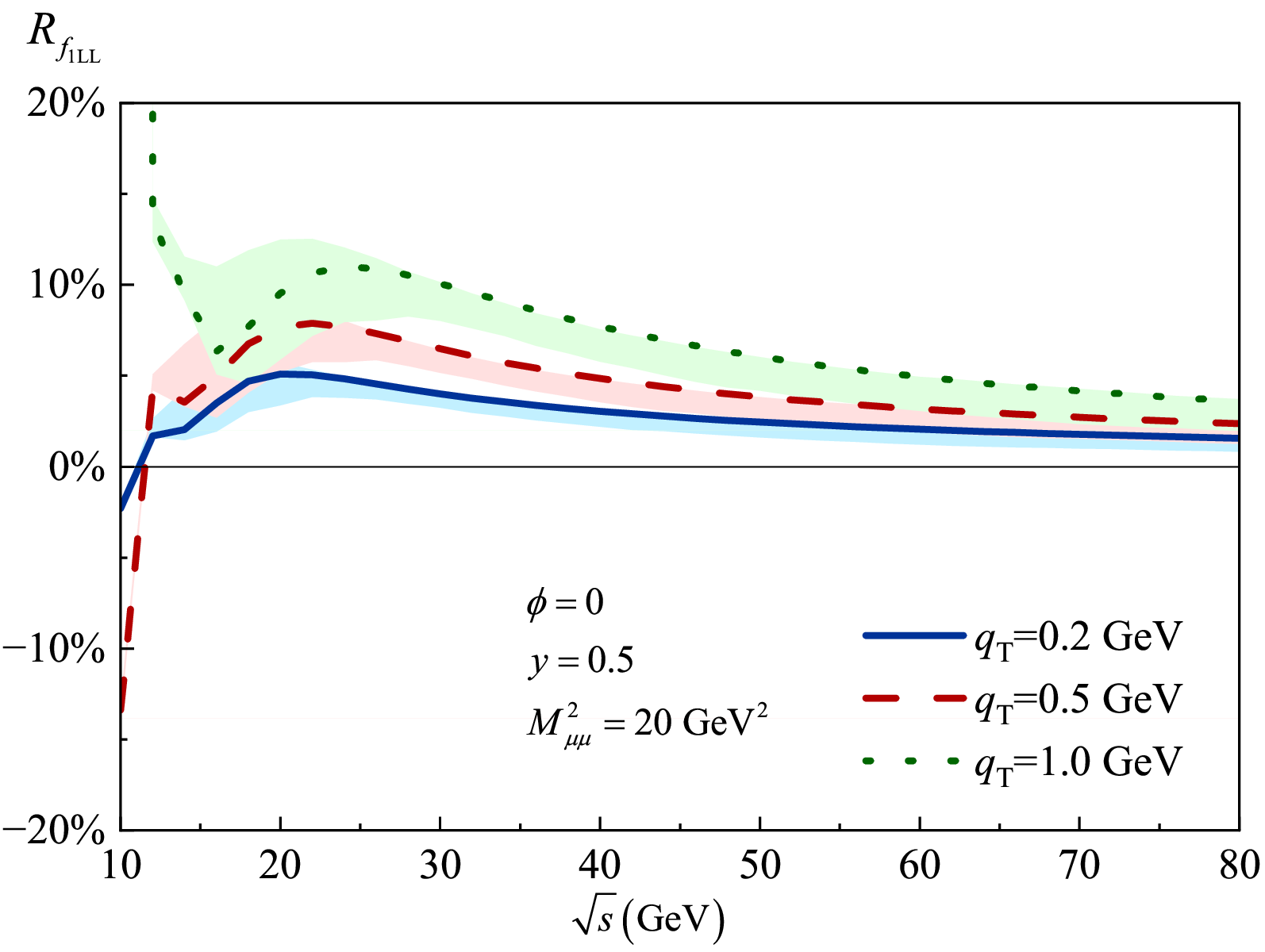}
\caption{Relative $f_{1LL}^g$ correction to $A_{E_{xy}}$ as a function of $\sqrt{s}$ for $Q^2=M_{\mu\mu}^2=20~\mathrm{GeV}^2$, $q_T=0.2$, $0.5$, and $1.0~\mathrm{GeV}$, with $y=0.5$ and $\phi=0$.}
\label{fig:f1LL-correction-s}
\end{figure}

The corresponding energy dependence is shown in Fig.~\ref{fig:f1LL-correction-s}. Around $\sqrt{s}=20$--$30~\mathrm{GeV}$, the correction is approximately $4$--$6\%$, $6$--$8\%$, and $10\%$ for $q_T=0.2$, $0.5$, and $1.0~\mathrm{GeV}$, respectively, and decreases to only a few percent at higher energies.

At very low $\sqrt{s}$, the allowed $x_b$ range is shifted toward the large-$x_b$ tail, making the relative correction increasingly sensitive to the suppressed $f_{1LL}^g$ contribution, the changing partonic-channel weights, numerical accuracy, and cancellations in the convolution. We therefore do not attach quantitative significance to the correction below approximately $\sqrt{s}=15~\mathrm{GeV}$. In the kinematic region where the spectator model is under reasonable control, the gluonic $f_{1LL}^{g/d}$ term gives a modest correction and leaves the main $Q^2$ and $\sqrt{s}$ dependence of $A_{E_{xy}}$ essentially unchanged.

\section{Conclusion}\label{section5}

We have investigated the proton-deuteron Drell-Yan phenomenology of the tensor-polarized double-helicity-flip gluon distribution $h_{1TT}^g$ using the low-scale input from our spin-$1$ spectator model. The distribution was evolved with its leading-order autonomous QCD equation for the full replica ensemble and then inserted into the Drell-Yan convolution. Compared with the earlier benchmark in which the unknown double-helicity-flip gluon distribution was modeled by a longitudinally polarized gluon PDF, our calculation generally predicts smaller cross sections and a much stronger $Q^2$ dependence of the asymmetry. The magnitude $|A_{E_{xy}}|$ falls rapidly with the hard scale, whereas its dependence on $\sqrt{s}$ becomes weak once the convolution is dominated by the small- and intermediate-$x$ region.

We also considered the gluonic $f_{1LL}^g$ contribution to the reference cross-section sum. Starting from the spectator-model gluon input, we evolved $f_{1LL}$ in the coupled singlet-gluon system and used only its gluon component in this auxiliary comparison. The resulting shift in $A_{E_{xy}}$ is modest over the kinematic region where the spectator model is reasonably stable and does not change the qualitative behavior associated with $h_{1TT}^g$.

The main limitation of the present calculation comes from kinematics that force the convolution into $x_b\gtrsim0.2$, where the spectator-model distributions are strongly suppressed and less stable. This affects the low-$\sqrt{s}$ and large-$Q^2$ regions in particular. Likewise, the very small-$q_T$ region lies outside the range in which the fixed-order treatment should be viewed as a precision description. Subject to these limitations, proton-deuteron Drell-Yan production remains a useful channel for testing the double-helicity-flip gluon structure of the deuteron, and the present calculation gives a model-based estimate that includes both QCD evolution and the uncertainty of the nonperturbative input.

\section*{Acknowledgements}
This work is partially supported by the National Natural Science Foundation of China under Grant No. 12150013. X. X. is also supported by the SEU Innovation Capability Enhancement Plan for Doctoral Students under Grant No. CXJH$\_$SEU 25138.


\begin{thebibliography}{99}
\bibitem{Hoodbhoy:1988am}
P.~Hoodbhoy, R.~L.~Jaffe and A.~Manohar,
Nucl. Phys. B \textbf{312} (1989), 571-588
doi:10.1016/0550-3213(89)90572-5

\bibitem{Bacchetta:2000jk}
A.~Bacchetta and P.~J.~Mulders,
Phys. Rev. D \textbf{62} (2000), 114004
doi:10.1103/PhysRevD.62.114004
[arXiv:hep-ph/0007120 [hep-ph]].

\bibitem{Boer:2016xqr}
D.~Boer, S.~Cotogno, T.~van Daal, P.~J.~Mulders, A.~Signori and Y.~J.~Zhou,
JHEP \textbf{10} (2016), 013
doi:10.1007/JHEP10(2016)013
[arXiv:1607.01654 [hep-ph]].

\bibitem{Cotogno:2017puy}
S.~Cotogno, T.~van Daal and P.~J.~Mulders,
JHEP \textbf{11} (2017), 185
doi:10.1007/JHEP11(2017)185
[arXiv:1709.07827 [hep-ph]].

\bibitem{Close:1990zw}
F.~E.~Close and S.~Kumano,
Phys. Rev. D \textbf{42} (1990), 2377-2379
doi:10.1103/PhysRevD.42.2377

\bibitem{Hino:1998ww}
S.~Hino and S.~Kumano,
Phys. Rev. D \textbf{59} (1999), 094026
doi:10.1103/PhysRevD.59.094026
[arXiv:hep-ph/9810425 [hep-ph]].

\bibitem{Hino:1999qi}
S.~Hino and S.~Kumano,
Phys. Rev. D \textbf{60} (1999), 054018
doi:10.1103/PhysRevD.60.054018
[arXiv:hep-ph/9902258 [hep-ph]].

\bibitem{Kumano:2010vz}
S.~Kumano,
Phys. Rev. D \textbf{82} (2010), 017501
doi:10.1103/PhysRevD.82.017501
[arXiv:1005.4524 [hep-ph]].

\bibitem{Kumano:2016ude}
S.~Kumano and Q.~T.~Song,
Phys. Rev. D \textbf{94} (2016) no.5, 054022
doi:10.1103/PhysRevD.94.054022
[arXiv:1606.03149 [hep-ph]].

\bibitem{HERMES:2005pon}
A.~Airapetian \textit{et al.} [HERMES],
Phys. Rev. Lett. \textbf{95} (2005), 242001
doi:10.1103/PhysRevLett.95.242001
[arXiv:hep-ex/0506018 [hep-ex]].

\bibitem{Khan:1991qk}
H.~Khan and P.~Hoodbhoy,
Phys. Rev. C \textbf{44} (1991), 1219-1222
doi:10.1103/PhysRevC.44.1219

\bibitem{Cosyn:2017fbo}
W.~Cosyn, Y.~B.~Dong, S.~Kumano and M.~Sargsian,
Phys. Rev. D \textbf{95} (2017) no.7, 074036
doi:10.1103/PhysRevD.95.074036
[arXiv:1702.05337 [hep-ph]].

\bibitem{Miller:2013hla}
G.~A.~Miller,
Phys. Rev. C \textbf{89} (2014) no.4, 045203
doi:10.1103/PhysRevC.89.045203
[arXiv:1311.4561 [nucl-th]].

\bibitem{Slifer:2014bda}
K.~Slifer,
J. Phys. Conf. Ser. \textbf{543} (2014) no.1, 012003
doi:10.1088/1742-6596/543/1/012003

\bibitem{Maxwell:2018gci}
J.~Maxwell, D.~Crabb, D.~Day, W.~Detmold, R.~Jaffe, M.~Jones, C.~Keith, D.~Keller, D.~Meekins and R.~Milner, \textit{et al.}
[arXiv:1803.11206 [nucl-ex]].

\bibitem{Keller:2022abm}
D.~Keller [SpinQuest],
[arXiv:2205.01249 [nucl-ex]].

\bibitem{Arbuzov:2020cqg}
A.~Arbuzov, A.~Bacchetta, M.~Butenschoen, F.~G.~Celiberto, U.~D'Alesio, M.~Deka, I.~Denisenko, M.~G.~Echevarria, A.~Efremov and N.~Y.~Ivanov, \textit{et al.}
Prog. Part. Nucl. Phys. \textbf{119} (2021), 103858
doi:10.1016/j.ppnp.2021.103858
[arXiv:2011.15005 [hep-ex]].

\bibitem{Aidala:2019pit}
C.~A.~Aidala, A.~Bacchetta, M.~Boglione, G.~Bozzi, V.~Carassiti, M.~Chiosso, R.~Cimino, G.~Ciullo, M.~Contalbrigo and U.~D'Alesio, \textit{et al.}
[arXiv:1901.08002 [hep-ex]].

\bibitem{Accardi:2012qut}
A.~Accardi, J.~L.~Albacete, M.~Anselmino, N.~Armesto, E.~C.~Aschenauer, A.~Bacchetta, D.~Boer, W.~K.~Brooks, T.~Burton and N.~B.~Chang, \textit{et al.}
Eur. Phys. J. A \textbf{52} (2016) no.9, 268
doi:10.1140/epja/i2016-16268-9
[arXiv:1212.1701 [nucl-ex]].

\bibitem{AbdulKhalek:2021gbh}
R.~Abdul Khalek, A.~Accardi, J.~Adam, D.~Adamiak, W.~Akers, M.~Albaladejo, A.~Al-bataineh, M.~G.~Alexeev, F.~Ameli and P.~Antonioli, \textit{et al.}
Nucl. Phys. A \textbf{1026} (2022), 122447
doi:10.1016/j.nuclphysa.2022.122447
[arXiv:2103.05419 [physics.ins-det]].

\bibitem{Anderle:2021wcy}
D.~P.~Anderle, V.~Bertone, X.~Cao, L.~Chang, N.~Chang, G.~Chen, X.~Chen, Z.~Chen, Z.~Cui and L.~Dai, \textit{et al.}
Front. Phys. (Beijing) \textbf{16} (2021) no.6, 64701
doi:10.1007/s11467-021-1062-0
[arXiv:2102.09222 [nucl-ex]].

\bibitem{Jaffe:1989xy}
R.~L.~Jaffe and A.~Manohar,
Phys. Lett. B \textbf{223} (1989), 218-224
doi:10.1016/0370-2693(89)90242-6

\bibitem{Artru:1989zv}
X.~Artru and M.~Mekhfi,
Z. Phys. C \textbf{45} (1990), 669
doi:10.1007/BF01556280

\bibitem{Kumano:2019igu}
S.~Kumano and Q.~T.~Song,
Phys. Rev. D \textbf{101} (2020) no.5, 054011
doi:10.1103/PhysRevD.101.054011
[arXiv:1910.12523 [hep-ph]].

\bibitem{Kumano:2020gfk}
S.~Kumano and Q.~T.~Song,
Phys. Rev. D \textbf{101} (2020) no.9, 094013
doi:10.1103/PhysRevD.101.094013
[arXiv:2003.06623 [hep-ph]].

\bibitem{Nzar:1992ax}
M.~Nzar and P.~Hoodbhoy,
Phys. Rev. D \textbf{45} (1992), 2264-2268
doi:10.1103/PhysRevD.45.2264

\bibitem{Detmold:2016gpy}
W.~Detmold and P.~E.~Shanahan,
Phys. Rev. D \textbf{94} (2016) no.1, 014507
[erratum: Phys. Rev. D \textbf{95} (2017) no.7, 079902]
doi:10.1103/PhysRevD.94.014507
[arXiv:1606.04505 [hep-lat]].

\bibitem{Mulders:2000sh}
P.~J.~Mulders and J.~Rodrigues,
Phys. Rev. D \textbf{63} (2001), 094021
doi:10.1103/PhysRevD.63.094021
[arXiv:hep-ph/0009343 [hep-ph]].

\bibitem{Jakob:1997wg}
R.~Jakob, P.~J.~Mulders and J.~Rodrigues,
Nucl. Phys. A \textbf{626} (1997), 937-965
doi:10.1016/S0375-9474(97)00588-5
[arXiv:hep-ph/9704335 [hep-ph]].

\bibitem{Bacchetta:2008af}
A.~Bacchetta, F.~Conti and M.~Radici,
Phys. Rev. D \textbf{78} (2008), 074010
doi:10.1103/PhysRevD.78.074010
[arXiv:0807.0323 [hep-ph]].

\bibitem{Meissner:2007rx}
S.~Meissner, A.~Metz and K.~Goeke,
Phys. Rev. D \textbf{76} (2007), 034002
doi:10.1103/PhysRevD.76.034002
[arXiv:hep-ph/0703176 [hep-ph]].

\bibitem{Lu:2016vqu}
Z.~Lu and B.~Q.~Ma,
Phys. Rev. D \textbf{94} (2016) no.9, 094022
doi:10.1103/PhysRevD.94.094022
[arXiv:1611.00125 [hep-ph]].

\bibitem{Bacchetta:2020vty}
A.~Bacchetta, F.~G.~Celiberto, M.~Radici and P.~Taels,
Eur. Phys. J. C \textbf{80} (2020) no.8, 733
doi:10.1140/epjc/s10052-020-8327-6
[arXiv:2005.02288 [hep-ph]].

\bibitem{Xie:2022lra}
X.~Xie and Z.~Lu,
Phys. Lett. B \textbf{842} (2023), 137973
doi:10.1016/j.physletb.2023.137973
[arXiv:2210.16532 [hep-ph]].

\bibitem{Xie:2024cnv}
X.~Xie and Z.~Lu,
Phys. Lett. B \textbf{856} (2024), 138961
doi:10.1016/j.physletb.2024.138961
[arXiv:2405.11474 [hep-ph]].

\bibitem{Bacchetta:2024fci}
A.~Bacchetta, F.~G.~Celiberto and M.~Radici,
Eur. Phys. J. C \textbf{84} (2024) no.6, 576
doi:10.1140/epjc/s10052-024-12927-y
[arXiv:2402.17556 [hep-ph]].

\bibitem{Xie:2026uja}
X.~Xie, D.~Y.~Chen and Z.~Lu,
Phys. Rev. D \textbf{114} (2026) no.3, 034003
doi:10.1103/h3ym-dkr1
[arXiv:2603.15224 [hep-ph]].

\bibitem{Kumano:2022cdj}
S.~Kumano and Q.~T.~Song,
Rev. Mex. Fis. Suppl. \textbf{3} (2022) no.3, 0308097
doi:10.31349/SuplRevMexFis.3.0308097
[arXiv:2201.04875 [hep-ph]].

\bibitem{AbdulKhalek:2019mzd}
R.~Abdul Khalek \textit{et al.} [NNPDF],
Eur. Phys. J. C \textbf{79} (2019) no.6, 471
doi:10.1140/epjc/s10052-019-6983-1
[arXiv:1904.00018 [hep-ph]].

\bibitem{NNPDF:2021njg}
R.~D.~Ball \textit{et al.} [NNPDF],
Eur. Phys. J. C \textbf{82} (2022) no.5, 428
doi:10.1140/epjc/s10052-022-10328-7
[arXiv:2109.02653 [hep-ph]].

\bibitem{Botje:2010ay}
M.~Botje,
Comput. Phys. Commun. \textbf{182} (2011), 490-532
doi:10.1016/j.cpc.2010.10.020
[arXiv:1005.1481 [hep-ph]].
\end{thebibliography}
\end{document}